\documentclass[aps,prb,reprint,preprintnumbers]{revtex4-2}
\usepackage{graphicx}
\usepackage{bm}
\usepackage{epsfig}
\usepackage{ulem}
\usepackage{color}
\usepackage{mathrsfs}
\usepackage{dcolumn}
\usepackage{setspace}
\usepackage{array}
\usepackage{amsmath}
\usepackage{amssymb}
\usepackage{dsfont}
\usepackage{gensymb}
\usepackage{dcolumn}
\usepackage{multirow}
\usepackage{bibentry,natbib}
\usepackage{booktabs}
\usepackage{appendix}
\usepackage{extarrows}
\usepackage[utf8]{inputenc}

\begin{document}
\bibliographystyle{unsrt}

\title{Searching Repulsive Casimir Forces Between Magneto-Electric Materials}
\author{Zixuan Dai$^1$,  Qing-Dong Jiang$^{1,2}$}
\email{qingdong.jiang@sjtu.edu.cn}
\affiliation{$^1$Tsung-Dao Lee Institute \& School of Physics and Astronomy, Shanghai Jiao Tong University, Pudong, Shanghai, 201210, China\\
$^2$Shanghai Branch, Hefei National Laboratory, Shanghai, 201315, China.
}

\begin{abstract}
The Casimir effect, arising from vacuum quantum fluctuations, plays a fundamental role in the development of modern quantum electrodynamics. In parallel, the field of condensed matter has flourished through the discovery of various materials exhibiting broken symmetries, often connected to topology and characterized by magneto-electric coupling. To enhance the comprehension of the role of parity symmetry and time-reversal symmetry in determining the sign of the Casimir force, we calculate the Casimir forces between magneto-electric materials and obtain a phase diagram governing the sign of symmetry-breaking-induced Casimir forces. We also investigate how the force phase diagram varies with the separation distances between the objects. Our results contribute to a better understanding of the sign of the Casimir force, a subject bearing both theoretical interest and practical significance.

\end{abstract}
\maketitle

%%%Introduction%%%
\section{\label{Introduction}Introduction} The Casimir effect is a macroscopic manifestation of quantum fluctuations, which predicts the attractive force between two parallel, charge-neutral metal plates due to the bigger zero-point pressure outside the plates \cite{Casimir1}. Lifshitz generalized the Casimir force formula to two dielectrics \cite{Lifshitz1956}, and Pitaevskii theoretically demonstrated that the repulsive Casimir force can be achieved by inserting medium 3 between materials 1,2 if their permittivities satisfy $-[\epsilon_1(i\omega)-\epsilon_3(i\omega)][\epsilon_2(i\omega)-\epsilon_3(i\omega)]>0$ \cite{Pitaevskii}. A few decades after the theoretical prediction, Lamoreaux confirmed the existence of a Casimir force using a torsion pendulum \cite{Lamoreaux}, and many other experiments have measured the Casimir force with high precision \cite{Munday2018, 2022sphere-plate-sphere, 2021gratings, 2011thermal, 2019AFMClean}. In particular, repulsive Casimir forces have been discovered in the laboratory between the gold sphere and the dielectric plate immersed in fluids such as bromobenzene or ferrofluids \cite{Munday2009, Zhang2024}. 

Repulsive Casimir forces are pursued for applications in micro/nanoelectromechanical systems \cite{2024MEMS}, vacuum levitation \cite{2014Levi,Levitation2015,Levitation2016} and superlubricity \cite{superlubricity2008}. However, achieving this goal is not straightforward.  \cite{Klichtheorem} proved a famous, yet discouraging, no-go theorem that the Casimir force between two reciprocal bodies related by parity symmetry is always attractive. To achieve a repulsive Casimir force, one may employ two objects, one with large dielectric function and the other with large magnetic permeability \cite{2023magnetodielectric, 2018magnetodielectric, 2014magnetodielectric, meta2008rosa, metamaterial2008}, or with complex geometries \cite{spheresphere2008, cylinderplate2008, sphereplate2013, needlehole2010, siliconnanostruc2017}, or with non-reciprocal materials \cite{TI2011, PTI2017, quantized2012, ChernInsulator2014, WSM2015, 2020chiralWSM, 2023nonrecip, 2024PEMC}. In 2019, the authors in \cite{Jiang2019} proposed a novel method to achieve repulsive Casimir forces by inserting a third parity-breaking material between two materials, offering a universal solution to achieving repulsive Casimir forces between materials with similar properties, including traditional metallic plates.

The no-go theorem establishes parity symmetry breaking as a necessary condition for realizing Casimir repulsion \cite{Klichtheorem}. However, this conclusion relies on specific assumptions—most notably, the reciprocity of the materials—which are not explicitly stated in the original formulation. Recent developments have revealed that repulsive Casimir forces can also arise between time-reversal symmetry-breaking materials, even in parity-symmetric configurations, as demonstrated in systems such as topological insulators, Chern insulators, quantum Hall states, and Weyl semimetals.
\cite{TI2011, PTI2017, quantized2012, ChernInsulator2014, WSM2015, 2020chiralWSM, TopologicalMatter2020}. Theoretical studies have also shown that the non-reciprocal components of materials can contribute repulsive terms to the Casimir force when two objects are mirror images of each other \cite{2022nonrecip}. In other words, Casimir repulsion can be achieved by breaking time-reversal symmetry while preserving parity symmetry. Despite numerous case-by-case investigations into Casimir-related phenomena, a comprehensive understanding of how discrete symmetry breaking influences the emergence of repulsive forces remains largely unexplored.

In this work, we aim to deepen the understanding of the connection between discrete symmetry breaking and Casimir forces by computing the Casimir interaction between magneto-electric materials that can simultaneously break time-reversal and parity symmetries. Specifically, we focus on the most general class of linear, isotropic media—bi-isotropic materials (BIM) \cite{Lindellbook, Silva2022BIM}. We obtain a phase diagram that maps the sign of the Casimir force as a function of the strength of parity and time-reversal symmetry breaking. By systematically exploring a broad parameter space, our results reveal that time-reversal symmetry breaking plays a more fundamental role than parity symmetry breaking in realizing repulsive Casimir forces. Due to the frequency dependence of the magneto-electric response functions, we further observe that the phase diagram evolves with separation distance, indicating that Casimir repulsion is more readily achieved at larger distances.

%%%Set-up and formula%%%
\section{Casimir Force and Bi-isotropic Materials}
\begin{figure}[!htb]
    \centering
    \includegraphics[width = 0.4\textwidth]{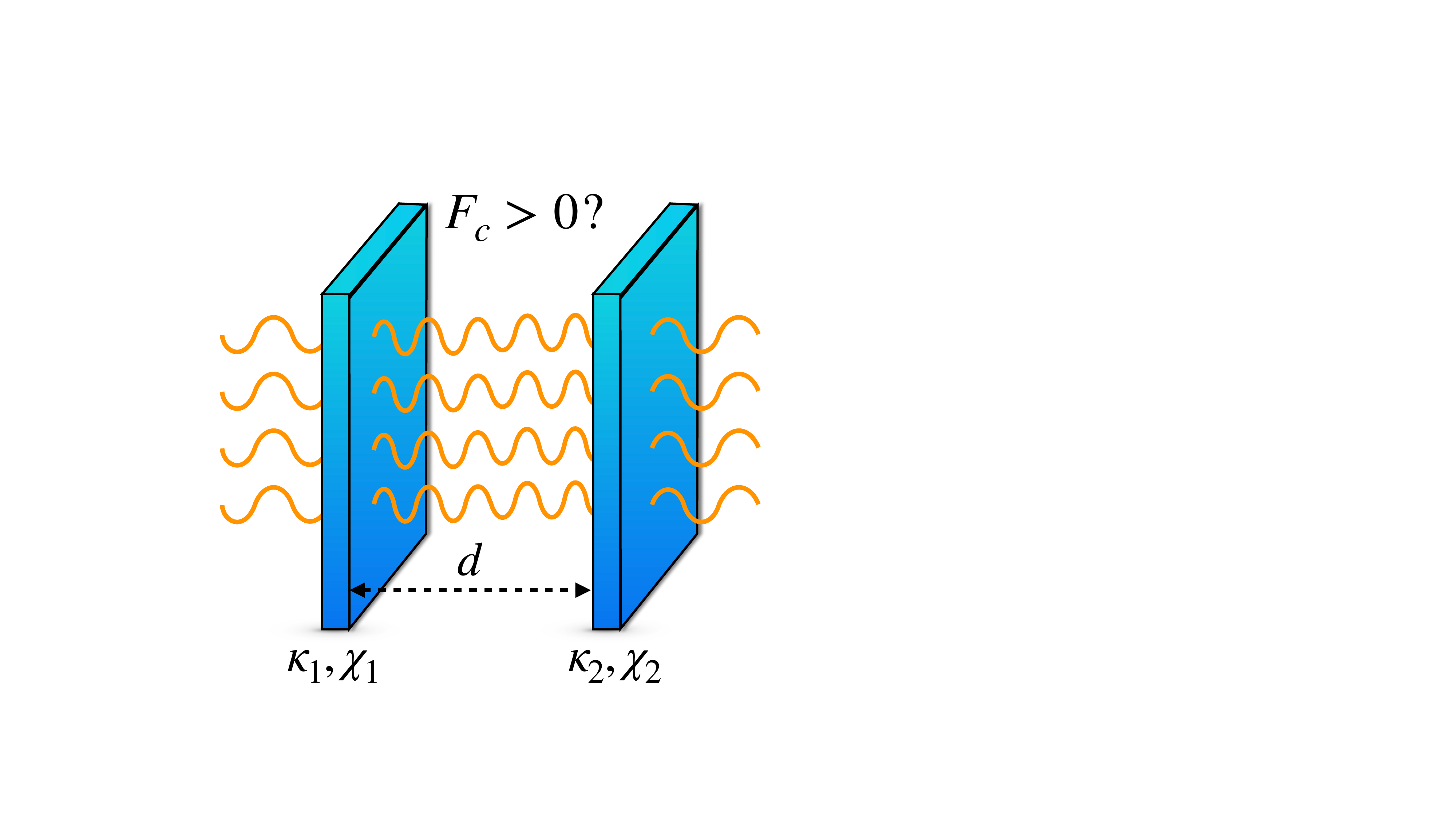}
    \caption{\label{fig:setup} Schematic diagram of Casimir effect between two bi-isotropic plates which are separated with distance $d$. Each BIM plate is characterized by four parameters: $\epsilon_i, \mu_i, \chi_i, \kappa_i(i = 1,2)$. The permittivity and permeability for two plates are assumed to be the same: $\epsilon_1 = \epsilon_2 = \epsilon, \mu_1 = \mu_2 = \mu$. } 
    \label{fig1}
\end{figure}

We investigate the Casimir forces between two parallel bi-isotropic plates separated by distance $d$ (Fig. \ref{fig:setup}). The electromagnetic response of the bi-isotropic plates is governed by the following constitutive relations: 
\begin{equation}
\begin{aligned}
\textbf{D} &=  \epsilon \textbf{E} + (\chi - i\kappa)\sqrt{\epsilon_0\mu_0}\textbf{H} \nonumber \\
\textbf{B} &=  \mu \textbf{H} + (\chi + i\kappa) \sqrt{\epsilon_0\mu_0} \textbf{E} 
\end{aligned}
\end{equation}
where $\epsilon$ is the permittivity and $\mu$ is the permeability. The magneto-electric coupling parameters $\chi$ and $\kappa$ are of essential importance here: $\chi$ and $\kappa$ indicate the non-reciprocity and chirality of the BIM, respectively. BIM with $\chi\neq0$ is non-reciprocal and breaks the restricted time reversal symmetry \cite{2020RTRS}; whereas BIM with $\kappa\neq0$ is chiral and breaks the parity symmetry. The bi-isotropic relation provide a unified description for diverse physical systems, ranging from natural materials like chiral molecules \cite{Condonmodel} and intrinsic magnetoelectrics \cite{1960Dzyaloshinski, 2005Fiebig} to engineered systems such as chiral metamaterials \cite{EMChirality2, 2020Chiroptical}, Tellegen metamaterials \cite{2024OpticalTellegen, 2025GiantTellegen}, topological insulators \cite{2009TI}, and is related to axion electrodynamics \cite{2023AxionF, 2023AxionS}. In artificial metamaterials, $\kappa$ and $\chi$ can be several orders larger than that of natural materials (typical $\sim10^{-4}$) and reach values on the order of 0.1, enabling strong and tunable bi-isotropic effects for controlling light-matter interactions. Here we consider the Casimir force between two plates sharing the same permittivity $\epsilon$ and permeability $\mu$, while having distinguished magneto-electric properties $\chi_i$, $\kappa_i$ ($i=1,2$).

The Casimir force between two parallel plates can be derived either through the stress-tensor approach (see Appendix A) \cite{PTI2017,2012Buhmann,2018ZPstresstensor} or via the path-integral formalism \cite{2009Kardar}, which we adopt here. Both methods yield the same result. At zero temperature, the force can be expressed in terms of reflection coefficients of the two plates: 
\begin{equation}\label{Ec}
F_c = \frac{\hbar A}{\pi}\int_0^\infty d\xi \int \frac{d^2\textbf{k}_\parallel}{(2\pi)^2} K\; \textup{Tr} \frac{\textbf{R}_1 \cdot \textbf{R}_2 e^{-2Kd}}{1-\textbf{R}_1 \cdot \textbf{R}_2 e^{-2Kd}}
\end{equation}
% E_c = \frac{\hbar A}{2\pi}\int_0^\infty d\xi \int \frac{d^2\textbf{k}_\parallel}{(2\pi)^2}\textup{ln\, det} (1-\textbf{R}_1 \cdot \textbf{R}_2 e^{-2Kd})
where $A$ is the plate area, $\textbf{k}_\parallel$ is the in-plane component of the wave vector, $\xi = -i\omega$ is the imaginary frequency, $K = \sqrt{\frac{\xi^2}{c^2}+{\textbf{k}^2_\parallel}}$. The $2\times2$ reflection matrix $\textbf{R}_i$ of plate $i$ ($i = 1,2$) takes the form
\begin{equation}
\textbf{R}_i = \left[
\begin{array}{cc}
r_{i,ss}(i\xi, \textbf{k}_\parallel) & r_{i,sp}(i\xi, \textbf{k}_\parallel)\\
r_{i,ps}(i\xi, \textbf{k}_\parallel) & r_{i,pp}(i\xi, \textbf{k}_\parallel)\\
\end{array}
\right]
\end{equation}
$r_{i,ps} (r_{i,sp})$ is the reflection coefficient from TE (TM) wave to TM (TE) wave of plate $i$. For clarity and subsequent analysis, we explicitly express the Casimir force in terms of the reflection coefficients as follows:
\begin{equation}\label{F2}
    F = -\frac{\hbar}{2\pi^2}\int_0^\infty d\xi \int_0^\infty k_\parallel dk_\parallel K \frac{Xe^{-2Kd} - 2Ye^{-4Kd}}{1-Xe^{-2Kd}+Ye^{-4Kd}}
\end{equation}
with 
\begin{equation}
\begin{aligned}
    X = & (r_{1,ss}r_{2,ss}+r_{1,pp}r_{2,pp}+r_{1,sp}r_{2,ps}+r_{1,ps}r_{2,sp}) \\
    Y = &(r_{1,ss}r_{1,pp}-r_{1,sp}r_{1,ps})(r_{2,ss}r_{2,pp}-r_{2,sp}r_{2,ps})
\end{aligned}   
\end{equation}
The reflection coefficients at the vacuum-BIM interface are determined by the material's electromagnetic properties and can be expressed as (see Appendix B for derivation):
\begin{footnotesize}
\begin{eqnarray}\label{coeff}
r_{ss,pp}&=&\frac{1}{\Delta}\left[2\eta_0\eta(c_0^2 - c_+c_-)\sqrt{1-(\frac{\chi}{n})^2}\pm(\eta^2 - \eta_0^2)c_0(c_+ + c_-)\right] \nonumber\\
r_{sp,ps}&=& \frac{2\eta_0\eta c_0}{\Delta}\left[\pm i(c_+ - c_-)\sqrt{1-(\frac{\chi}{n})^2} - (c_+ + c_-)\frac{\chi}{n}\right]
\end{eqnarray}
\end{footnotesize}
where the impedance $\eta_0 = \sqrt{{\mu_0}/{\epsilon_0}}$, $\eta = \sqrt{{\mu}/{\epsilon}}$. 
$c_0 = \cos\theta_0 = \sqrt{k_0^2-\textbf{k}^2_\parallel}/{k_0}$, where $\theta_0$ is the incident angle and $k_0 = \omega/c$.
$c_{\pm} = \cos\theta_{\pm} = {\sqrt{k^2_{\pm}-\textbf{k}^2_\parallel}}/{k_{\pm}}$, where $\theta_{\pm}$ is the refracted angles and $k_{\pm} = k_0(\sqrt{n^2-\chi^2}\pm\kappa)$ with the refraction index $n = \sqrt{\epsilon\mu/\epsilon_0\mu_0}$
. $\Delta = (\eta^2 + \eta_0^2)c_0(c_+ + c_-) + 2\eta_0\eta(c_0^2 + c_+c_-)\sqrt{1-({\chi}/{n})^2}$. 
In the absence of magneto-electric coupling, Eq.(\ref{coeff}) recovers the pure magnetodielectric case, yielding the well-known reflection coefficients: $r_{ss} = \frac{\eta c_0 - \eta_0 c_n}{\eta c_0 + \eta_0 c_n}$, $r_{pp} = \frac{\eta_0 c_0 - \eta c_n}{\eta_0 c_0 + \eta c_n}$, $r_{sp}=r_{ps}=0$, where $c_n = c_+ = c_-$.

For a dielectric-based metamaterial exhibiting bi-isotropic response, the resonant behaviors can be modeled in the following relations:
\begin{eqnarray}\label{dispersion}
\epsilon_r(\omega) &=& 1 + \frac{\omega_p^2}{\omega_R^2 - \omega^2 - i\omega\gamma(\omega)}\nonumber \\
\mu_r(\omega) &=& 1 + \frac{\omega_m^2}{\omega_R^2 - \omega^2 - i\omega\gamma(\omega)} + B + \frac{B\omega^2}{\omega_R^2 - \omega^2 - i\omega\gamma(\omega)}\nonumber \\
\kappa(\omega) &=& \pm\frac{\omega_\kappa \omega}{\omega_R^2 - \omega^2 - i\omega\gamma(\omega)}\nonumber \\
\chi(\omega) &=& \pm\frac{\omega_{\chi}^2}{\omega_R^2 - \omega^2 - i\omega\gamma(\omega)}
\end{eqnarray}
where $\epsilon_r = \epsilon/\epsilon_0$ and $\mu_r = \mu/\mu_0$. The dielectric function's dispersion relation is analogous to the Lorentz model characterized by the plasma frequency $\omega_p$, the resonant frequency $\omega_R$, but with a frequency-dependent dissipation parameter $\gamma(\omega)$. 
% The first two terms in this dispersion model describe the intrinsic magnetic response and the last two terms describe the effective magnetic response coming from the chirality of the materials. 
The permeability's dispersion relation includes a Lorentzian term characterized by $\omega_m$, and geometry-dependent terms parameterized by $B$ encoding the metamaterial's internal structural configuration (e.g., split-ring resonator dimensions \cite{ChiralDisp1999} or helix structure \cite{chiralmeta2010}). For the chiral parameter, we adopt the Condon model where $\omega_\kappa$ quantifies the resonant strength of the chirality \cite{Condonmodel, EMChirality2}. The sign is determined by the structural handedness, e.g. the plus (minus) sign corresponds to metamaterials composed of left-handed (right-handed) helical elements \cite{chiralmeta2010}, which is tunable via fabrication. For the non-reciprocal response, we model the frequency dependence using a Lorentzian form, where $\omega_\chi$ quantifies the resonant strength of non-reciprocity. This choice is supported by  \cite{2024OpticalTellegen}, where the Tellegen response in fabricated optical Tellegen metamaterials was shown to obey Lorentz-model behavior. In  \cite{2024OpticalTellegen}, the sign of the Tellegen metamaterial is decided by the saturation magnetization direction of the constituent ferromagnetic nanocylinder. We emphasize that the plus-minus sign for $\kappa(\omega)$ and $\chi(\omega)$ is crucial, as it has notable implications for the sign of the Casimir force. For mirror-symmetric plates: (i) reciprocal terms produce attraction, while (ii) non-reciprocal terms induce repulsion \cite{2022nonrecip}. Therefore, we expext that repulsive Casimir force are favored when plates have identical signs for $\kappa$ and opposite signs for $\chi$, as this configuration maximizes repulsive non-reciprocal interactions while circumventing the parity-symmetry-enforced attractive contributions from reciprocal coupling.

Real material systems are subject to several physical constraints, like passivity to ensure positive energy dissipation, and sum rules governing integrated spectral responses. Specifically, the passivity of the BIM requires $\omega^2_p\omega^2_m\geq\omega^4_{\chi}$ and $B\omega^2_p\geq\omega^2_{\kappa}$ (see Appendix \ref{app:PassiveCondition}), which impose critical bounds on the achievable magneto-electric coupling strengths.
For the sum rules, taking the dielectric function as an example, the bounded moment $\int^\infty_{-\infty}\frac{d\omega}{\pi}\omega^n\frac{\rm{Im}\epsilon_r(\omega)}{\omega}$ for arbitrary integer $n\geq0$ is required. This implies that $\gamma(\omega)$ must decay faster than any power of $\omega$ at high frequencies \cite{Chaikin_Lubensky_1995}. Clearly, a constant damping factor fails to provide the correct description of high-frequency behavior. However, since the Casimir force is predominantly determined by the material responses in the low-frequency range $\xi\in[0, \frac{c}{d}]$ (see section \ref{Phase Diagram}). Therefore, as long as the dispersion model accurately captures the low-frequency behavior, the computed Casimir force remains robust. In our dispersion model, we extend beyond the conventional constant damping approximation by selecting $\gamma(\omega) = \gamma\frac{1}{1-i\lambda\omega}$ so that the dispersion model satisfies the sum rule up to $n = 4$ \cite{Chaikin_Lubensky_1995}. We choose the phenomenological parameter $\lambda\omega_R\ll1$ so that the variation of $\gamma(\omega)$ is small below the resonant frequency.

\section{\label{Phase Diagram}Phase Diagram for the Sign of Casimir Force}
To identify the optimal parameter combinations for achieving Casimir repulsion, we first investigate the Casimir interaction between plates with asymmetric magneto-electric properties. We vary the chirality and non-reciprocal parameters of one plate while maintaining the parameters of the other plate unchanged. For convenience, we use positive (negative) $\omega_{\kappa_i}/\omega_{\chi_i}(i = 1,2)$ to represent the plus (minus) sign in the dispersion. We set the resonant frequency $\omega_{R} = 10^{15}rad/s$, which is practicable in the metamaterial fabrication. To maximize the magneto-electric effects mediated by $\kappa, \chi$ and keep the contribution from $\epsilon$ and $\mu$ as small as possible, we select parameters $\omega_{R} = \omega_p = \omega_m, B = 0.04$, where passivity of BIM imposes the constraints $|\omega_{\kappa_i}/\omega_R|\leq 0.2, |\omega_{\chi_i}/\omega_R|\leq 1 $. For plate 1, we choose $\omega_{\kappa_1}/\omega_R = 0.2, \omega_{\chi_1}/ \omega_R = 1$. For plate 2, we tune the parameters $\omega_{\kappa_2}, \omega_{\chi_2}$ across the entire range permitted by passivity and see how the Casimir force changes.

\begin{figure}[!htb]
    \centering
    \includegraphics[width = 0.5\textwidth]{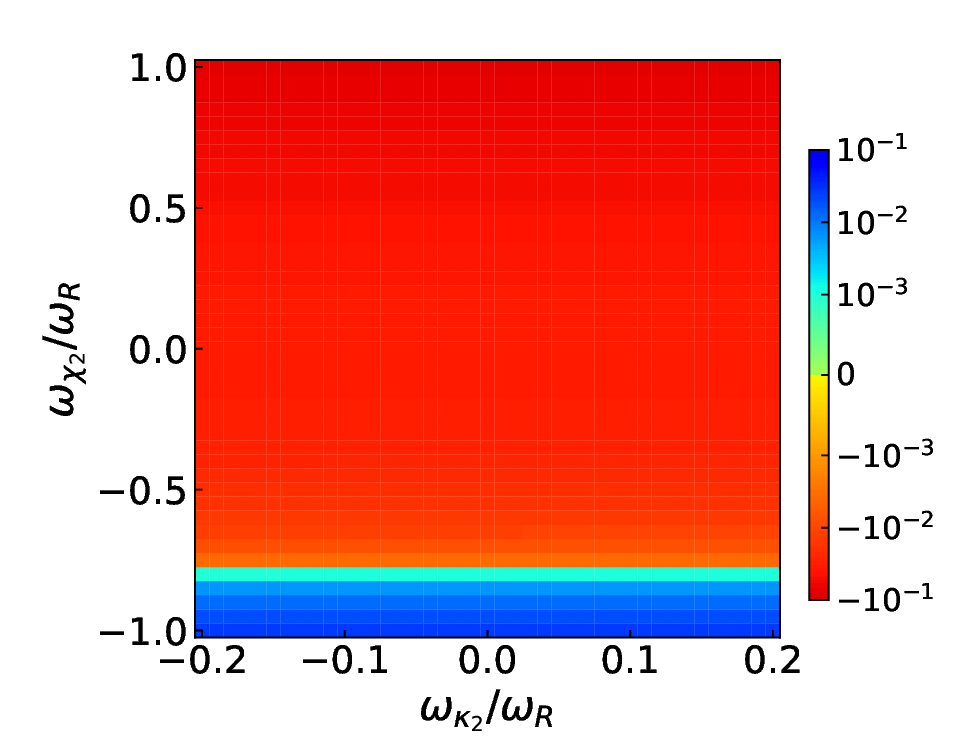}
    \caption{The phase diagram of the Casimir force $F_c/F_0$ between two BIM plates with separation $d = 1\mu m$. The reference force $F_0 = \frac{\pi^2\hbar c A}{240d^4}$ is the magnitude of the Casimir force between two parallel perfect metallic plates. The red (blue) region represents attractive (repulsive) force. $\omega_{\kappa_1} = 0.2\omega_R, \omega_{\chi_1} = \omega_R$ is fixed and the Casimir force varies with different $\omega_{\kappa_2}, \omega_{\chi_2}$. The sign of the $\omega_{\kappa_i}, \omega_{\chi_i}$ represents the sign in the dispersion model of $\kappa_i, \chi_i$. The parameters for the permittivity and permeability are: $\omega_{R} = \omega_p = \omega_m = 10^{15} rad/s, B = 0.04, \gamma = 0.05\omega_R, \lambda = 10^{-5}\omega_R^{-1}$, which impose the constraints $|\omega_{\kappa_i}/\omega_R|\leq 0.2, |\omega_{\chi_i}/\omega_R|\leq 1 $.} 
    \label{fig:fix1side}
\end{figure}

\begin{figure}[!htb]
    \centering
    \includegraphics[width = 0.5\textwidth]{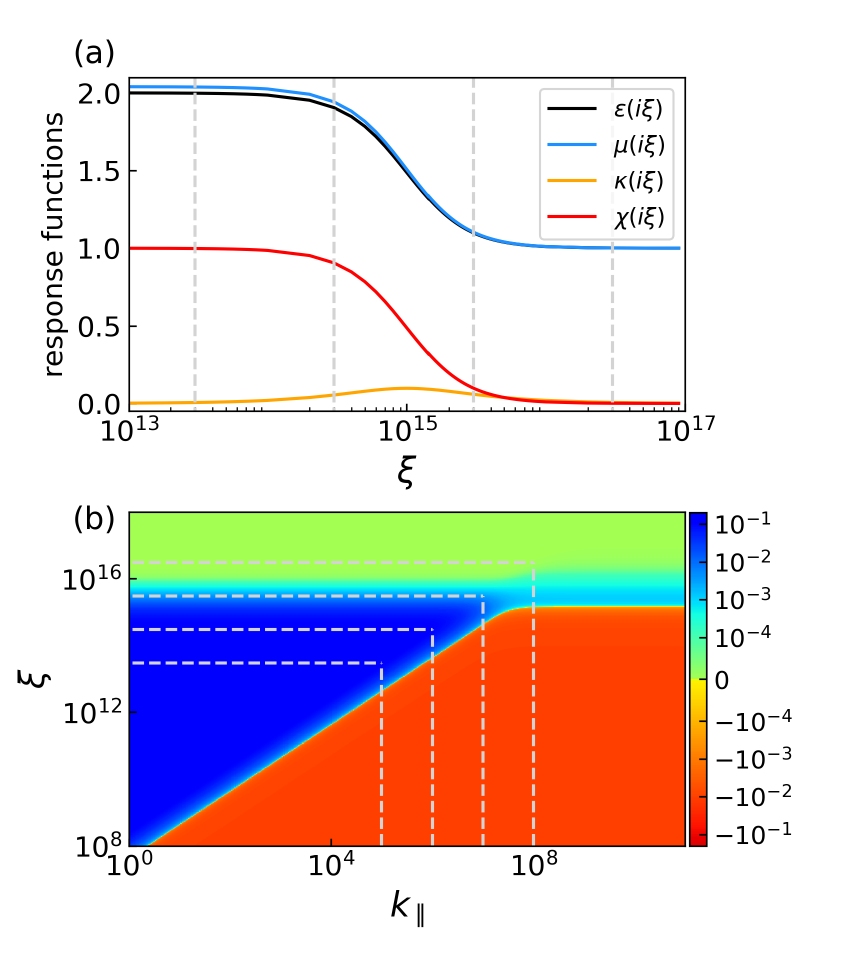}
    \caption{The response functions and ($-r^2_{ss}-r^2_{pp}+r^2_{ps}+r^2_{sp}$) for $\omega_\kappa = 0.2\omega_R, \omega_\chi = \omega_R$. (a) Dispersion of the material response functions $\epsilon(i\xi)$, $\mu(i\xi)$, $\chi(i\xi)$ and $\kappa(i\xi)$. The vertical gray lines mark the characteristic frequency $\xi = \frac{c}{d}$ for separations $d = 10, 1, 0.1, 0.01 \mu m$ (left to right). $\epsilon(i\xi)$, $\mu(i\xi)$, $\chi(i\xi)$ are real and decrease monotonically. $\kappa(i\xi)$ is purely imaginary.  
    (b) The reflection coefficient combination ($-r^2_{ss}-r^2_{pp}+r^2_{ps}+r^2_{sp}$) as a function of $k_\parallel$ and $\xi$. The horizontal and vertical gray lines corresponds to the characteristic frequency $\xi = \frac{c}{d}$ and cut-off momentum $k_\parallel = \frac{1}{d}$ for the same set of separations. Red (blue) regions indicate negative (positive) values, corresponding to attractive (repulsive) contributions to the Casimir force.
    Parameters for permittivity and permeability are the same as those in Fig.\ref{fig:fix1side}.}
    \label{fig:dispersion and spectrum}
\end{figure}

The result is shown in Fig.\ref{fig:fix1side}. Both attractive (red region) and repulsive (blue region) Casimir force can be achieved. However, achieving repulsive Casimir force is hard and requires very strong non-reciprocal responses. The weak dependence on the chirality parameter originates from two key factors, the narrow spectral region of Condon model dispersion relation at imaginary frequency and the small bounded value allowed by passivity (see Fig.\ref{fig:dispersion and spectrum}(a)).
In our case, the Casimir force is repulsive when the two BIM plates are in parity-symmetric configuration ($\omega_{\kappa_2}/\omega_R = -0.2, \omega_{\chi_2}/\omega_R = -1$), where the no-go theorem \cite{Klichtheorem} does not apply since the objects are non-reciprocal. Here, the repulsion arising from the non-reciprocal part dominates over the attraction from the reciprocal part, resulting in a net repulsive force. This highlights the critical importance of distinguishing between reciprocal and non-reciprocal responses when analyzing Casimir interactions in mirror-image setups. On the other hand, the phase diagram shows that the best configuration for achieving repulsion occurs when $\omega_{\kappa_2}/\omega_R = 0.2, \omega_{\chi_2}/\omega_R = -1$, consistent with the theoretical predictions in Section 2 that repulsion favors identical $\kappa$ signs and opposite $\chi$ signs. 
By properly tuning the parameters, one can control both the sign and the magnitude of the Casimir force. 
Different from the case of topological insulators \cite{TI2011} and chiral metamaterials \cite{Zhao2009}, now we consider two tuning parameters with different symmetry properties simultaneously, which broadens the scope for exploring repulsive Casimir forces. 

\begin{figure}[!htb]
\centering
\includegraphics[width = 0.55\textwidth]{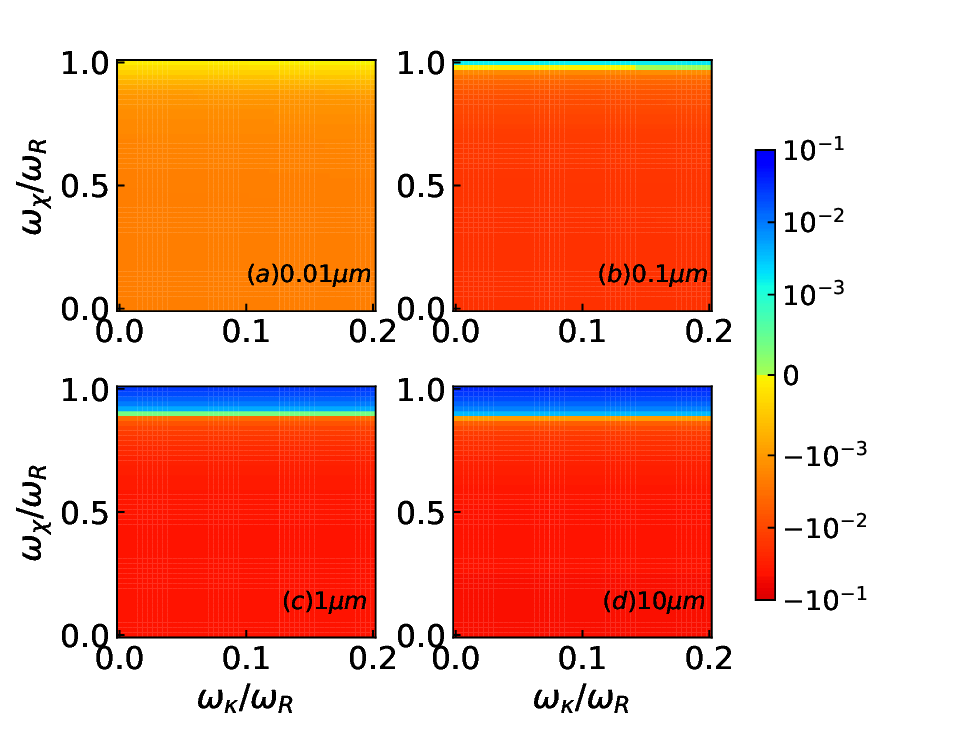}
\caption{\label{fig:PDvsd}Phase diagram of the Casimir force $F_c(d)/F_0(d)$ at varying distances: (a) $d = 0.01\mu m$, (b) $d = 0.1\mu m$, (c) $d = 1\mu m$, (d) $d = 10\mu m$. The reference force $F_0(d) = \frac{\pi^2 \hbar c A}{240d^4}$ is the magnitude of the force between two perfect conducting plates at each distance. The parameters are chosen such that $\omega_{\chi1} = -\omega_{\chi2} = \omega_{\chi}, \omega_{\kappa1}= \omega_{\kappa2} = \omega_{\kappa}$. For fixed $\omega_{\kappa}, \omega_{\chi}$, Casimir repulsion emerges more readily at larger separations. Parameters in the permittivity and permeability are the same as those in Fig.\ref{fig:fix1side}.}
\end{figure}

%keypoint2
Drawing upon the insights obtained from the above results, we consider the configuration where $\omega_{\chi_1}=-\omega_{\chi_2}=\omega_{\chi}, \omega_{\kappa_1}=\omega_{\kappa_2}=\omega_{\kappa}$ in the following to obtain the force as repulsive as possible. In this configuration, we investigate how the Casimir force phase diagram evolves with the separation distance. Fig.\ref{fig:PDvsd} shows that as the distance between the two plates decreases, achieving repulsion becomes increasingly challenging. We note that when $\omega_{\chi} = \omega_{\kappa} = 0$, the two BIM plates reduce to identical magnetodielectric plates, resulting in an inevitably attractive force. When $\omega_{\chi} = 0$, the system reverts to the case of two chiral materials and the force is attractive for all the $\omega_{\kappa}$ allowed by the passive condition, consistent with previous research \cite{Zhao2009, CommentChiral2010, Restriction2010}. When $\omega_{\kappa} = 0$, the system reduces to the case of two Tellegen materials, enabling the generation of repulsive forces.

To gain a deeper understanding of the distance dependency of the force phase diagram, let's examine the general force formula Eq.(\ref{F2}). When $\omega_{\chi_1}=-\omega_{\chi_2}, \omega_{\kappa_1}=\omega_{\kappa_2}$, the reflection coefficients of two plates are related: $r_{1,ss} = r_{2,ss} = r_{ss}, r_{1,pp} = r_{2,pp} = r_{pp}, r_{1,sp} = -r_{2,ps} = r_{sp}, r_{1,ps} = -r_{2,sp} = r_{ps}$. The force formula can be simplified as 
\begin{equation}\label{I}
    F = -\frac{\hbar}{2\pi^2}\int_0^\infty d\xi \int_0^\infty k_\parallel dk_\parallel K \frac{xe^{-2Kd} - 2ye^{-4Kd}}{1-xe^{-2Kd}+ye^{-4Kd}}
\end{equation}
with 
\begin{equation}
\begin{aligned}
    x = & (r_{ss}^2+r_{pp}^2-r_{sp}^2-r_{sp}^2) \\
    y = &(r_{ss}r_{pp}-r_{sp}r_{ps})^2
\end{aligned}   
\end{equation}
where $K=\sqrt{{\frac{\xi}{c}}^2 + k_{\parallel}^2}$ and $r_{ss,pp,ps,sp}(i\xi, k_{\parallel})$ are reflection coefficients. Since the denominator of the integrand is always positive (see the proof in Appendix \ref{app:denominator}), the force's sign is determined by the numerator. While the numerator comprises of two terms, the first term generally dominates, and the sign of the Casimir force is predominantly determined by ($r^2_{ss}+r^2_{pp}-r^2_{ps}-r^2_{sp}$). 
The exponential factors in Eq.\eqref{I} suggest that the primary contribution to the integral comes from the low-frequency range $\xi \in [0, \frac{c}{d}]$. The distance sets a characteristic frequency below which the contribution dominates. Therefore, achieving a repulsive Casimir force requires maximizing ($-r^2_{ss}-r^2_{pp}+r^2_{ps}+r^2_{sp}$) when $\xi \in [0, \frac{c}{d}]$, which is attainable in BIM with large off-diagonal reflection coefficients owing to magneto-electric coupling. 
In addition, the resonant frequency $\omega_R$ is another characteristic frequency. If $\xi \gg \omega_R$, the responses of the BIM tend to vanish and $\epsilon(i\xi), \mu(i\xi)\rightarrow1$, $\chi(i\xi), \kappa(i\xi)\rightarrow0$. If $\xi \ll \omega_R$, the responses are almost constants identical to the static responses.
Comparing the two characteristic frequencies, we can understand why Casimir repulsion occurs more easily at larger distances.

The dispersion model and the quantity ($-r^2_{ss}-r^2_{pp}+r^2_{ps}+r^2_{sp}$) for specific parameters $\omega_\kappa = 0.2\omega_R, \omega_\chi = \omega_R$ are plotted in Fig.\ref{fig:dispersion and spectrum}. When $d = 1\mu m$ or $d = 10\mu m$, $\frac{c}{d} < \omega_R$ and response functions behave approximately as constants in the frequency range that mainly contributes (see Fig.\ref{fig:dispersion and spectrum}(a)). In these cases, ($-r^2_{ss}-r^2_{pp}+r^2_{ps}+r^2_{sp}$) have more positive regions than negative regions in the main contribution range $\xi\in[0, \frac{c}{d}]$, $k_\parallel\in[0, \frac{1}{d}]$ and we get a repulsive Casimir force after integrating over $\xi$ and $k_{\parallel}$. 
When $d = 0.1\mu m$, $\frac{c}{d} \sim \omega_R$ and the system begins to probe the resonant structure of the material response. In this case, since $\kappa(i\xi)$ is small, the Casimir force is primarily governed by $\epsilon(i\xi), \mu(i\xi)$ and $\chi(i\xi)$. The response functions $\epsilon(i\xi), \mu(i\xi), \chi(i\xi)$ monotonically decrease with $\xi$. The smaller response functions near $\frac{c}{d}$ result in a less positive ($-r^2_{ss}-r^2_{pp}+r^2_{ps}+r^2_{sp}$) when $\xi \sim \frac{c}{d}$. In the end, we get a less repulsive Casimir force after integration. 
When the separation is further reduced to $d = 0.01\mu m$, the characteristic frequency satisfies $\frac{c}{d} \gg \omega_R$.
At such high frequencies (i.e., $\xi\in[0, \frac{c}{d}]$, $k_\parallel\in[0, \frac{1}{d}]$), the BIM response functions become negligible, and the integrand
($-r^2_{ss}-r^2_{pp}+r^2_{ps}+r^2_{sp}$) acquires more negative than positive contributions over the relevant integration domain. These observations explain why achieving repulsion becomes increasingly difficult as the plate separation decreases: the material response is less effective at high frequencies, and the contribution from repulsive components diminishes accordingly.

%\caption{\textcolor{red}{The response functions and ($-r^2_{ss}-r^2_{pp}+r^2_{ps}+r^2_{sp}$) for $\omega_\kappa = 0.2\omega_R, \omega_\chi = \omega_R$. (a) Dispersion of the material response functions $\epsilon(i\xi)$, $\mu(i\xi)$, $\chi(i\xi)$ and $\kappa(i\xi)$. The vertical gray lines mark the characteristic frequency $\xi = \frac{c}{d}$ for separations $d = 10, 1, 0.1, 0.01 \mu m$ (left to right). $\epsilon(i\xi)$, $\mu(i\xi)$, $\chi(i\xi)$ are real and decrease monotonically with $\xi$, exhibiting static (vanishing) response at low (high) frequency $\xi\ll\omega_R$ ($\xi\gg\omega_R$). In contrast, $\kappa(i\xi)$ is purely imaginary, with its dominant response concentrated near the resonant frequency.  (b) The reflection coefficient combination ($-r^2_{ss}-r^2_{pp}+r^2_{ps}+r^2_{sp}$) as a function of parallel wave vector $k_\parallel$ and frequency $\xi$. The horizontal and vertical grey lines corresponds to the characteristic frequency $\xi = \frac{c}{d}$ and cut-off momentum $k_\parallel = \frac{1}{d}$ for the same set of separations. Red (blue) regions indicate negative (positive) values, corresponding to attractive (repulsive) contributions to the Casimir force.Parameters for permittivity and permeability are the same as those in Fig.\ref{fig:fix1side}.}}

\begin{figure}[!htb]
    \centering
    \includegraphics[width = 0.5\textwidth]{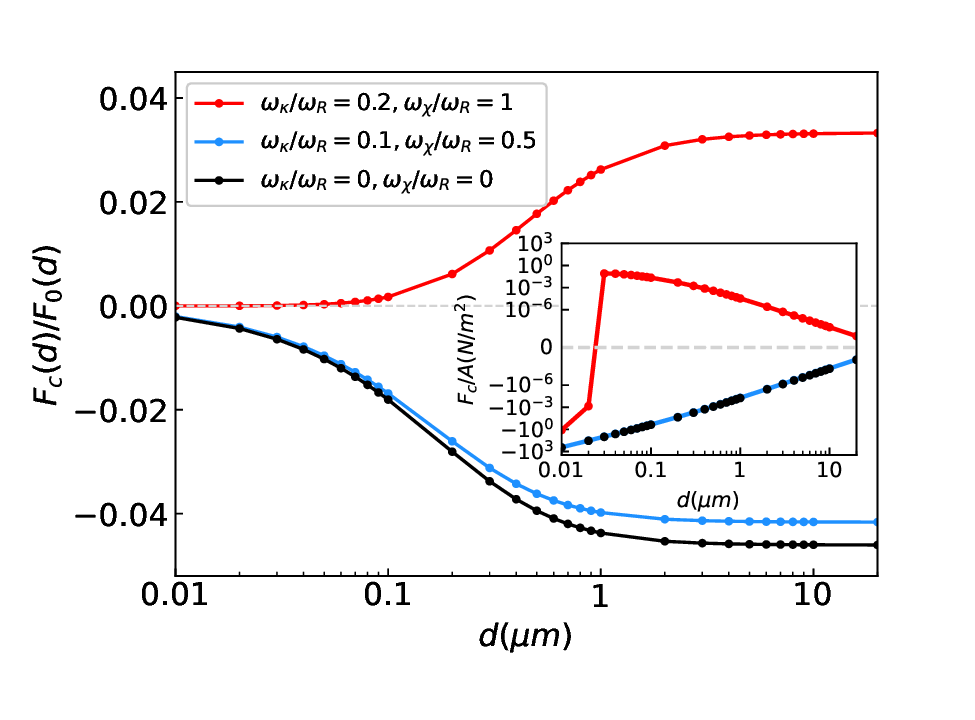}
    \caption{\label{fig:Fvsd}Distance dependence of the normalized Casimir force $F_c(d)/F_0(d)$ between two BIM plates. The magneto-electric coupling parameters are chosen such that $\omega_{\chi_1}=-\omega_{\chi_2}=\omega_{\chi}, \omega_{\kappa_1}=\omega_{\kappa_2}=\omega_{\kappa}$, with other parameters matching Fig.\ref{fig:fix1side}. 
    When $\omega_\kappa/\omega_R=0.2,\omega_\chi/\omega_R=1$ (red curve), the force is attractive at short distance but repulsive at large distance. When $\omega_\kappa/\omega_R=\omega_\chi/\omega_R=0$ (black curve) or $\omega_\kappa/\omega_R=0.1,\omega_\chi/\omega_R=0.5$ (blue curve), the force is always attractive.
    The inset figure shows the total force per unit area: $F_c(d)/A$. In the case of $\omega_\kappa/\omega_R = \omega_\chi/\omega_R = 0$ (black dots) and $\omega_\kappa/\omega_R = 0.1, \omega_\chi/\omega_R = 0.5$ (blue curve), the attractive force are nearly identical and decay with the distance. }
\end{figure}

% keypoint3
In Fig.\ref{fig:Fvsd}, we calculate the Casimir forces versus the distance for different $\omega_\chi$ and $\omega_\kappa$, considering separation range which covers both the retarded and non-retarded limit. We obtained two kinds of distance-dependent force behaviors: long-range attraction or initial attraction followed by repulsion at greater distances. These behaviors are recognizable from changes in the phase diagram with distance. For example, we can see from Fig.\ref{fig:PDvsd} that when $\omega_{\kappa} = 0.2\omega_R$ and $\omega_{\chi} = \omega_R$, the force undergoes a sign reversal as the separation distance varies. From the phase diagram of the Casimir force, we can also see that the sign-changing distance decreases as $\omega_{\chi}$ or $\omega_{\kappa}$ becomes larger. Since $\frac{c}{d}>10\omega_R$ and $\frac{c}{d}<0.1\omega_R$ corresponds to the non-retarded and retarded limit and the sign reversal point must lie in between, we estimate that $d_c\sim\frac{c}{\omega_R}$, e.g. $d_c\sim 0.3~{\rm \mu m}$ in our case. When $\omega_\chi = 0, \omega_\kappa= 0$, the Casimir force recovers the case of two magneto-dielectric plates. In the retarded limit, the force scales with $d^{-4}$, hence the normalized force is approximately constant. In the non-retarded limit, the scaling depends on the electric and magnetic properties of each plate \cite{2012Buhmann}. 
% What's more, we do not find any equilibrium position between two BIM plates standing in vacuum.  

%%%Summary%%%
\section{Conclusion}
We calculate the Casimir force between two BIM plates, showing that it is possible to get both repulsive and attractive force by adjusting the non-reciprocal parameters and chirality parameters. Through systematically
exploring a broad parameter space, we obtain the force phase diagrams and show that time-reversal symmetry breaking plays a more fundamental role than parity symmetry breaking in realizing repulsive Casimir forces.
Furthermore, by examining how the phase diagram evolves with the plate separation, we show that repulsion is more readily achieved at larger distances. This behavior is explained by comparing two characteristic frequencies: the cutoff frequency $\frac{c}{d}$ , determined by the plate separation, and the material-specific resonance frequency $\omega_R$. We show two distinct types of force–distance relationships: one exhibiting long-range attraction, and another characterized by a transition from attraction at short distances to repulsion at larger separations. 

Finally, we emphasize that beyond establishing the phase diagram of repulsive Casimir forces for materials with $\chi \neq 0$ and $\kappa \neq 0$, our theoretical framework provides a unified approach for calculating Casimir forces in magneto-electric systems, bridging the gap between chiral and nonreciprocal materials, which are often treated separately. Our analysis further reveals that the Casimir force between bi-isotropic plates becomes repulsive once the magnetoelectric coupling parameters $\kappa$ and $\chi$ surpass critical thresholds, offering a promising strategy to mitigate stiction in nanoscale devices. Even outside this repulsive regime, when the force remains attractive, its magnitude is still markedly reduced compared to that between conventional metallic plates, offering a means to mitigate adhesive problem in nano-devices. There are already many experimental examples of materials with isolated chiral responses (e.g. chiral molecules, chiral metamaterials) or non-reciprocal responses (e.g. intrinsic magnetoelectric material Cr$_2$O$_3$, topological insulator, Tellegen metamaterial). 
To achieve a metamaterials with both non-vanishing $\kappa$ and $\chi$, one can in principle design a bi-isotropic material by randomly distributing randomly oriented chiral components and Tellegen components in the material. Alternatively, we can construct a material with bi-isotropic responses on the boundary by putting a thin chiral material on a Tellegen material, or vice versa. 

\bigskip
The work is supported by National Natural Science Foundation of China (NSFC) under Grant No.12374332 and Innovation Program for Quantum Science and Technology Grant No.2021ZD0301900, and Shanghai Science and Technology Innovation Action Plan Grant No. 24LZ1400800.

\appendix
\section{Stress-Tensor Approach for Deriving Casimir Force Between BIM Plates}\label{app:StressTensor}
Using the stress-tensor formalism, the expression of the Casimir force between two parallel bi-isotropic plates standing in vacuum can be written as \cite{PTI2017}:
\begin{eqnarray}
    \mathbf{F} = -\frac{\hbar}{2\pi}\int_0^{\infty}d\xi\int_{\partial V}d\mathbf{A} \cdot \left\{\left[\frac{\xi^2}{c^2}\mathbf{G}(\mathbf{r},\mathbf{r}',i\xi) \right.\right. \nonumber\\ 
    \left.\left. + \frac{\xi^2}{c^2}\mathbf{G}^{T}(\mathbf{r}',\mathbf{r},i\xi) + \overrightarrow{\nabla} \times \mathbf{G}(\mathbf{r},\mathbf{r}',i\xi) \times \overleftarrow{\nabla}{'} \right.\right. \nonumber \\
    \left.\left. + \overrightarrow{\nabla} \times \mathbf{G}^{T}(\mathbf{r}',\mathbf{r},i\xi) \times \overleftarrow{\nabla}{'} \right]
    - \textup{Tr}\left[\frac{\xi^2}{c^2} \mathbf{G}(\mathbf{r},\mathbf{r}',i\xi) \right.\right. \nonumber \\
    \left.\left. + \overrightarrow{\nabla} \times \mathbf{G}(\mathbf{r},\mathbf{r}',i\xi) \times \overleftarrow{\nabla}{'}\right]\right\}_{\mathbf{r}'\rightarrow\mathbf{r}} 
\end{eqnarray}
with the Green's function satisfies the equation
\begin{equation}\label{Greenfunc}
    \left[\nabla\times\frac{1}{\mu_r}\nabla\times + \frac{\xi^2}{c^2} \epsilon_r - \hat V(\chi,\kappa)
    \right]\mathbf{G}(\mathbf{r}, \mathbf{r}', i\xi) = \mathbf{I}\delta (\bold r-\bold r^{\prime})
\end{equation}
where all contributions from $\chi$ and $\kappa$ are encoded in the operator $\hat V$: $$\hat V(\chi,\kappa)\equiv \frac{\xi}{c}\frac{\chi-i\kappa}{\mu_r}\nabla\times - \frac{\xi}{c}\nabla\times\frac{\chi+i\kappa}{\mu_r} + \frac{\xi^2}{c^2}\frac{\chi^2+\kappa^2}{\mu_r}$$
According to perturbation theory, $\mathbf{G}=\mathbf{G}_0+\mathbf{G}_0 
\hat V(\chi, \kappa)\mathbf{G}_0 +\dots$, where $\mathbf{G}_0$ denotes the Green's function for $\chi=\kappa=0$ materials. And $\mathbf{G}^{T}$ can be expanded in the same way. 

With the stress-tensor formalism, the contributions from $\chi$ and $\kappa$ are evident. 
For small values of $\chi$ and $\kappa$, one may readily separate their effects from the normal contribution by retaining only the leading terms in the Green's function expansion. 
However, in this work we consider the case of large $\chi$ and $\kappa$, which are moreover frequency- and spatially dependent, making such a decomposition impractical. Instead, we use the exact reflection coefficients of BIM plates to calculate the Casimir force (see Eq.(\ref{Ec})).

\section{Reflection Coefficients of BIM}\label{app:ReflectCoeff}
In the \textbf{E}, \textbf{H} representation, the constitutive relations for BIM can be written as 
\begin{equation} \label{D1}
\begin{aligned} 
\textbf{D} &= \epsilon \textbf{E} + (\chi - i\kappa) \sqrt{\epsilon_0\mu_0}\textbf{H} \\ 
\textbf{B} &= \mu \textbf{H} + (\chi + i\kappa) \sqrt{\epsilon_0\mu_0} \textbf{E}
\end{aligned}
\end{equation}
\noindent
For later convenience, we rewrite the the constitutive relations for BIM in the \textbf{E}, \textbf{B} representation
\begin{equation}\label{D2}
\begin{aligned} 
\textbf{D} &=  \tilde{\epsilon} \textbf{E} + \alpha \textbf{B} \\  
\textbf{H} &=  \frac{1}{\mu} \textbf{B} + \beta \textbf{E} 
\end{aligned}
\end{equation}
with $\tilde{\epsilon} = \epsilon - \frac{1}{\mu}(\chi^2 + \kappa^2)\epsilon_0\mu_0$, $\alpha = \frac{1}{\mu}(\chi - j\kappa)\sqrt{\epsilon_0\mu_0}$, $\beta = - \frac{1}{\mu}(\chi + j\kappa)\sqrt{\epsilon_0\mu_0}$. 
Considering time-harmonic electromagnetic plane waves $e^{i\textbf{k}\cdot\textbf{x}-i\omega t}$ and using Eq.(\ref{D2}), we have
\begin{equation} \label{D6}
\frac{1}{\mu\omega}\textbf{k}\times(\textbf{k}\times\textbf{E}) + \omega\tilde{\epsilon}\textbf{E} + (\alpha + \beta)(\textbf{k}\times\textbf{E}) = 0
\end{equation}
Since $\nabla\cdot\textbf{D} = \nabla\cdot(\tilde{\epsilon} \textbf{E} + \alpha \textbf{B}) = 0$ and $\nabla\cdot\textbf{B} = 0$, we have $\nabla\cdot\textbf{E} = 0$, i.e. $\textbf{k}\cdot\textbf{E} = 0$. Therefore, $\textbf{k}\times(\textbf{k}\times\textbf{E}) = (\textbf{k}\cdot\textbf{E})\textbf{k} - k^2\textbf{E} = -k^2\textbf{E}$ and Eq.(\ref{D6}) becomes 
\begin{equation} \label{D7}
(\frac{-k^2}{\mu\omega} + \omega\tilde{\epsilon})\textbf{E} + (\alpha + \beta)\textbf{k}\times\textbf{E} = 0
\end{equation}
or $M_{ij}E_j = 0$, where $M_{ij}$ is a tensor
\begin{equation} \label{D9}
M_{ij} = \left( 
\begin{array}{ccc}
    \frac{-k^2}{\mu\omega} + \omega\tilde{\epsilon} & -(\alpha + \beta)k_3 & (\alpha + \beta)k_2 \\
    (\alpha + \beta)k_3 & \frac{-k^2}{\mu\omega} + \omega\tilde{\epsilon} & -(\alpha + \beta)k_1 \\
    -(\alpha + \beta)k_2 & (\alpha + \beta)k_1 & \frac{-k^2}{\mu\omega} + \omega\tilde{\epsilon}
\end{array}
\right)
\end{equation}
For non-zero propagating modes, we require $\textup{det} M_{ij} = 0$. Then we have
\begin{equation} \label{D10}
\frac{-k^2}{\mu\omega} + \omega\tilde{\epsilon} = 0, \, \textup{or}\, (\frac{-k^2}{\mu\omega} + \omega\tilde{\epsilon})^2 = - (\alpha + \beta)^2k^2
\end{equation}
If $\frac{-k^2}{\mu\omega} + \omega\tilde{\epsilon} = 0$, then the non-zero propagating mode of $M_{ij}E_j = 0$ is $\textbf{E} = \textup{E}\hat{\textbf{k}}$. Since $\textbf{k}\cdot\textbf{E} = 0$, this solution is not allowed. If $(\frac{-k^2}{\mu\omega} + \omega\tilde{\epsilon})^2 = - (\alpha + \beta)^2k^2$, we have
\begin{equation} \label{D11}
\frac{k^2}{\omega^2\mu_0\epsilon_0} = \mu_r\epsilon_r - \chi^2 + \kappa^2 \pm 2|\kappa|\sqrt{\mu_r\epsilon_r - \chi^2}
\end{equation}
where $\mu_r$,$\epsilon_r$ are relative permittivity and relative permeability, respectively. Taking square root,
\begin{equation} 
\frac{k}{\omega\sqrt{\mu_0\epsilon_0}} = \pm(\sqrt{\mu_r\epsilon_r - \chi^2} \pm |\kappa|)
\end{equation}
When $\kappa\rightarrow 0, \chi \rightarrow 0$, the wave vector \textbf{k} should have normal dispersion relation $\frac{k}{\omega} = \sqrt{\mu\epsilon}$, so
\begin{equation} \label{D13}
\frac{k}{\omega\sqrt{\mu_0\epsilon_0}} = (\sqrt{\mu_r\epsilon_r - \chi^2} \pm |\kappa|)
\end{equation}\\
We define
\begin{equation} \label{D14}
\frac{k_{\pm}}{\omega\sqrt{\mu_0\epsilon_0}} = (\sqrt{\mu_r\epsilon_r - \chi^2} \pm \kappa)
\end{equation}\\
Next, we go on to find the propagating modes corresponding to the wave vectors $\textbf{k}_{\pm}$. Using (\ref{D14}), we have
\begin{align*}
\frac{-k_{\pm}^2}{\mu\omega} + \omega\tilde{\epsilon} 
& = \mp \frac{2\omega}{\mu} \kappa(\sqrt{\mu_r\epsilon_r - \chi^2} \pm \kappa)\epsilon_0\mu_0 \\
& = \mp\frac{2}{\mu}\kappa\sqrt{\epsilon_0\mu_0}k_{\pm} \\
& = (\alpha + \beta)(\pm jk_{\pm})
\end{align*}
So $M_{ij}$ becomes
\begin{equation} \label{D15}
M_{ij} = (\alpha + \beta)\left( 
\begin{array}{ccc}
    \pm jk_{\pm} & -k_3 & k_2 \\
    k_3 & \pm jk_{\pm} & -k_1 \\
    -k_2 & k_1 & \pm jk_{\pm}
\end{array}
\right)
\end{equation}
Solving out $M_{ij}E_j = 0$, we find the propagating modes corresponding to the wave vectors $\textbf{k}_{\pm}$ are 
\begin{equation} \label{D16}
\begin{aligned}
\boldsymbol{\epsilon_+} = 
\left(
\begin{array}{c}
k_3k_+ + jk_1k_2\\
-j(k_1^2 + k_3^2)\\
jk_2k_3 - k_1k_+
\end{array}
\right), 
\boldsymbol{\epsilon_-} = 
\left(
\begin{array}{c}
k_3k_- - jk_1k_2\\
j(k_1^2 + k_3^2)\\
-jk_2k_3 - k_1k_-
\end{array}
\right)
\end{aligned}
\end{equation}
where $N_{\pm}$ are normalization factors.

With the wave vector $\textbf{k}$ and propagating modes $\boldsymbol{\epsilon}_{\pm}$, the reflection coefficients for electromagnetic waves incident onto the vacuum-BIM interface can be found. Suppose the incident wave is in the x-z plane, then
\begin{align}
\textbf{E}_i & = \left[A_{\perp}\hat{y} + \frac{A_{\parallel}}{\omega\epsilon_0}(k_z\hat{x}-k_x\hat{z})\right]e^{i(k_x x + k_z z -\omega t)}\\
\textbf{H}_i & = \left[- \frac{A_{\perp}}{\mu_0\omega}(k_z\hat{x}-k_x\hat{z}) + A_{\parallel}\right]\hat{y}e^{i(k_x x + k_z z -\omega t)}
\end{align}
where $A_{\perp}(A_{\parallel})$ represents the TE(TM) wave. Similarly, the reflected wave has the form
\begin{align}
\textbf{E}_r & = \left[R_{\perp}\hat{y} - \frac{R_{\parallel}}{\omega\epsilon_0}(k_z\hat{x}+k_x\hat{z})\right]e^{i(k_x x - k_z z -\omega t)}\\
\textbf{H}_r & = \left[\frac{R_{\perp}}{\mu_0\omega}(k_z\hat{x}+k_x\hat{z}) + R_{\parallel}\hat{y}\right]e^{i(k_x x - k_z z -\omega t)}
\end{align}
Since $k_y = 0$, according to Eq.(\ref{D16}), the propagating modes are 
\begin{align} 
\boldsymbol{\epsilon}_+  = \left(
\begin{array}{c}
k_{3+}k_+\\ -jk_+^2\\ - k_{1+}k_+
\end{array} \right), 
\boldsymbol{\epsilon}_- = \left(
\begin{array}{c}
k_{3-}k_-\\ jk_-^2\\ - k_{1-}k_-
\end{array} \right)
\end{align}
Defining $s_{\pm} = \frac{k_x}{k_{\pm}}$, $c_{\pm} = \frac{k_{3\pm}}{k_{\pm}} = \frac{\sqrt{k_{\pm}^2 - k_x^2}}{k_{\pm}}$, the propagating modes can be written as
\begin{align} 
\boldsymbol{\epsilon}_+  = \left(
\begin{array}{c}
c_+\\ -j\\ - s_+
\end{array} \right), 
\boldsymbol{\epsilon}_-  = \left(
\begin{array}{c}
c_-\\ j\\ - s_-
\end{array} \right)
\end{align}
Besides, using Eq.(\ref{D1}), Eq.(\ref{D6}) and $\textbf{k}\times\textbf{E} = \omega\textbf{B}$, we have
\begin{equation}
\begin{aligned}\label{D17}
\textbf{E}_+ &= -j\eta_+\textbf{H}_+ \\ 
\textbf{E}_- &= j\eta_-\textbf{H}_- 
\end{aligned}  
\end{equation}
where
\begin{equation} \label{D19}
\eta_{\pm} = \sqrt{\frac{\mu}{\epsilon}}[\sqrt{1-(\frac{\chi}{n})^2} \mp j(\frac{\chi}{n})]
\end{equation}
$n = \sqrt{\frac{\epsilon\mu}{\epsilon_0\mu_0}}$ is the refraction index. Therefore, the transmitted wave have the form
\begin{gather}
\textbf{E}_t  = e_+ \left(
\begin{array}{c}
c_+\\ -j\\ - s_+
\end{array} \right) + e_- \left(
\begin{array}{c}
c_-\\ j\\ - s_-
\end{array} \right)
\\
\textbf{H}_t  = h_+ \left(
\begin{array}{c}
c_+\\ -j\\ - s_+
\end{array} \right) + h_-\left(
\begin{array}{c}
c_-\\ j\\ - s_-
\end{array} \right)
\end{gather}
Applying the boundary conditions at the vacuum-BIM interface, we have
\begin{equation}\label{D20}
\begin{aligned} 
\frac{A_{\parallel}}{\omega\epsilon_0}k_z - \frac{R_{\parallel}}{\omega\epsilon_0}k_z &= (-j\eta_+h_+)c_+ + (j\eta_+h_-)c_- \\
A_{\perp} + R_{\perp} &= -j(-j\eta_+h_+) + j(j\eta_+h_-) \\
- \frac{A_{\perp}}{\mu_0\omega}k_z + \frac{R_{\perp}}{\mu_0\omega}k_z &= h_+c_+ + h_-c_- \\
A_{\parallel} + R_{\parallel} &= -jh_+ + jh_-
\end{aligned} 
\end{equation}
Solving out Eq.(\ref{D20}), we obtain the reflection coefficients
\begin{small}
\begin{equation}
\begin{aligned}
r_{ss} &= \frac{1}{\Delta}[2\eta_0\eta(c_0^2 - c_+c_-)\sqrt{1-(\frac{\chi}{n})^2}+(\eta^2 - \eta_0^2)c_0(c_+ + c_-)]\\
r_{pp} &= \frac{1}{\Delta}[2\eta_0\eta(c_0^2 - c_+c_-)\sqrt{1-(\frac{\chi}{n})^2} - (\eta^2 - \eta_0^2)c_0(c_+ + c_-)] \\
r_{sp} &= \frac{2\eta_0\eta c_0}{\Delta}[i(c_+ - c_-)\sqrt{1-(\frac{\chi}{n})^2} - (c_+ + c_-)\frac{\chi}{n}] \\
r_{ps} &= -\frac{2\eta_0\eta c_0}{\Delta}[i(c_+ - c_-)\sqrt{1-(\frac{\chi}{n})^2} + (c_+ + c_-)\frac{\chi}{n}] 
\end{aligned}
\end{equation}
\end{small}
where $r_{ps}$($r_{sp}$) represents reflection coefficients from TE(TM) wave to TM(TE) wave, the denominator $\Delta = (\eta^2 + \eta_0^2)c_0(c_+ + c_-) + 2\eta_0\eta(c_0^2 + c_+c_-)\sqrt{1-(\frac{\chi}{n})^2}$, $\eta_0=\sqrt{\frac{\mu_0}{\epsilon_0}}, \eta=\sqrt{\frac{\mu}{\epsilon}}$. To obtain the reflection coefficients, $\eta_+\eta_- = \eta^2$, $\eta_+ + \eta_- = 2\eta\sqrt{1-(\frac{\chi}{n})^2}$ are used.

\section{Passive Conditions for Bi-isotropic Material}\label{app:PassiveCondition}
The BIM are characterized by the constitutive relations:
\begin{equation}
\begin{aligned} 
    \textbf{D} &= \epsilon_0\epsilon_r \textbf{E} + (\chi - i\kappa) \sqrt{\epsilon_0\mu_0}\textbf{H} \\ 
    \textbf{B} &= \mu_0\mu_r \textbf{H} + (\chi + i\kappa) \sqrt{\epsilon_0\mu_0} \textbf{E}
\end{aligned}
\end{equation}
For monochromatic electromagnetic fields with time dependence $\mathrm{exp}(-i\omega t)$, the average heating rate in a period per unit volume is \cite{LaudauEM}: 
\begin{equation}
    q = \frac{1}{2} \mathrm{Re} \{-i\omega(\textbf{E}^*\cdot\textbf{D} + \textbf{H}^*\cdot\textbf{B})\} 
\end{equation}
For BIM, 
\begin{equation}\label{q}
    q = \frac{\omega}{2}\left(\begin{array}{c}
    \textbf{E}^* \\ \textbf{H}^*
    \end{array}\right)\cdot\textbf{M}\cdot \left(\begin{array}{c}
    \textbf{E} \\ \textbf{H}
    \end{array}\right)
\end{equation}
with
\begin{equation}
\footnotesize
    \textbf{M} = \left(\begin{array}{cc}
    \mathrm{Im}\{\epsilon_0\epsilon_r\} & (\mathrm{Im}\{\chi\} + i\mathrm{Im}\{\kappa\})\sqrt{\epsilon_0\mu_0} \\
    (\mathrm{Im}\{\chi\} - i\mathrm{Im}\{\kappa\})\sqrt{\epsilon_0\mu_0} & \mathrm{Im}\{\mu_0\mu_r\}
    \end{array} \right)
\end{equation}
For passive materials, $q>0$ is required, i.e. the matrix $\textbf{M}$ should be positive definite. Therefore, $\mathrm{Im}\{\epsilon\}>0$, $\mathrm{Im}\{\mu\}>0$, $\mathrm{Im}\{\epsilon\}\cdot\mathrm{Im}\{\mu\} > (\mathrm{Im}\{\chi\})^2 + (\mathrm{Im}\{\kappa\})^2$. Applying the dispersion model introduced in Eq.(\ref{dispersion}), the passivity of BIM requires $\omega^2_p\omega^2_m\geq\omega^4_{\chi}$ and $B\omega^2_p\geq\omega^2_{\kappa}$. 

\section{The Proof that the Denominator in Eq.(\ref{I}) is Positive}\label{app:denominator}
In this section, we prove that the denominator in Eq.(\ref{I}) is always positive. The denominator in Eq.~(\ref{I}) can be rewritten as

\begin{eqnarray}
% &&1 - (r_{ss}^2+r_{pp}^2-r^2_{ps}-r^2_{sp})e^{-2Kd} \\
% &&\qquad + (r_{ss}r_{pp} - r_{sp}r_{ps})^2e^{-4Kd} \nonumber\\
% &=& 
&&[1-(r^2_{ss}+r^2_{pp})e^{-2Kd}+r^2_{ss}r^2_{pp}e^{-4Kd}] \\
&& \qquad+ [(r^2_{ps}+r^2_{sp})e^{-2Kd}-2r_{ss}r_{pp}r_{sp}r_{ps}e^{-4Kd}] \nonumber \\
&& \qquad
+ r^2_{sp}r^2_{ps}e^{-4Kd} \nonumber\\
&=& (1-r^2_{ss}e^{-2Kd})(1-r^2_{pp}e^{-2Kd}) \\
&& \qquad+(r^2_{ps}+r^2_{sp}-2r_{ss}r_{pp}r_{sp}r_{ps}e^{-2Kd})e^{-2Kd} \nonumber \\
&& \qquad
+ r^2_{sp}r^2_{ps}e^{-4Kd} \nonumber
\end{eqnarray}   
Since $r_{ss}(i\xi), r_{pp}(i\xi), r_{ps}(i\xi), r_{sp}(i\xi)$ are real and their absolute values $|r_{ss}(i\xi)|, |r_{pp}(i\xi)|, |r_{ps}(i\xi)|, |r_{sp}(i\xi)|\leq1$, the first term and the third term are positive. % (when $\chi^2 + \kappa^2 \leq n^2$)
Because $|r_{ss}r_{pp}e^{-2Kd}|<1$, the second term $(r^2_{ps}+r^2_{sp}-2r_{ss}r_{pp}r_{sp}r_{ps}e^{-2Kd})e^{-2Kd}>(r^2_{ps}+r^2_{sp}-2|r_{sp}r_{ps}|)e^{-2Kd}$ and it is also positive. Therefore, the denominator in Eq.~(\ref{I}) is always positive. 
% $|r_{ss}(i\xi)|, |r_{pp}(i\xi)|, |r_{ps}(i\xi)|, |r_{sp}(i\xi)|\leq1$

\normalem
\bibliographystyle{iopart-num} % We choose the "" reference style
\bibliography{ref} % Entries are in the refs.bib file

\end{document}